\documentclass[apj]{emulateapj}

\usepackage{apjfonts}
\usepackage{natbib}
\usepackage{amssymb, amsmath}
\usepackage{etoolbox}

\usepackage[
bookmarks=true,           
bookmarksnumbered=true,   
colorlinks=true,          
citecolor=blue,           
linkcolor=blue,           
menucolor=blue,           
urlcolor=blue,            
linkbordercolor={0 0 1},  
pdfborder={0 0 1},
frenchlinks=true]{hyperref}

\renewcommand{\eprint}[1]{\href{http://arxiv.org/abs/#1}{#1}}

\providecommand{\adsurl}[1]{\href{#1}{ADS}}

\makeatletter
\patchcmd{\NAT@citex}
  {\@citea\NAT@hyper@{%
     \NAT@nmfmt{\NAT@nm}%
     \hyper@natlinkbreak{\NAT@aysep\NAT@spacechar}{\@citeb\@extra@b@citeb}%
     \NAT@date}}
  {\@citea\NAT@nmfmt{\NAT@nm}%
   \NAT@aysep\NAT@spacechar\NAT@hyper@{\NAT@date}}{}{}

\patchcmd{\NAT@citex}
  {\@citea\NAT@hyper@{%
     \NAT@nmfmt{\NAT@nm}%
     \hyper@natlinkbreak{\NAT@spacechar\NAT@@open\if*#1*\else#1\NAT@spacechar\fi}%
       {\@citeb\@extra@b@citeb}%
     \NAT@date}}
  {\@citea\NAT@nmfmt{\NAT@nm}%
   \NAT@spacechar\NAT@@open\if*#1*\else#1\NAT@spacechar\fi\NAT@hyper@{\NAT@date}}
  {}{}
\makeatother

\makeatletter
\DeclareRobustCommand{\lowcase}[1]{\@lowcase#1\@nil}
\def\@lowcase#1\@nil{\if\relax#1\relax\else\MakeLowercase{#1}\fi}
\makeatother

\newcommand\degree{\degr}
\newcommand\degrees{\degree}
\newcommand\vs{{\em vs.}\ }
\newcommand\tnm[1]{\tablenotemark{#1}}
\newcommand\SST{{\em Spitzer Space Telescope}}
\newcommand\Spitzer{{\em Spitzer}}
\newcommand\chisq{$\chi^2$}
\newcommand\PT{$P$--$T$}

\newcommand\waspo{{wa008bs11}}
\newcommand\waspt[1]{{wa008bs2{#1}}}
\newcommand\waspf[1]{{wa008bs4{#1}}}
\newcommand\mnp{\emph{mnp}}

\newcommand\Teq{$T$\sb{eq}}
\newcommand\jhauth[1]{{#1}}
\newcommand\jhstud[1]{{#1}}

\DeclareSymbolFont{UPM}{U}{eur}{m}{n}
\DeclareMathSymbol{\umu}{0}{UPM}{"16}
\let\oldumu=\umu
\renewcommand\umu{\ifmmode\oldumu\else$\oldumu$\fi}
\newcommand\micro{\umu}
\renewcommand\micron{\micro m}
\newcommand\microns {\micron}
\newcommand\mctc{\multicolumn{2}{c}}
\let\oldsim=\sim
\renewcommand\sim{\ifmmode\oldsim\else$\oldsim$\fi}
\let\oldpm=\pm
\renewcommand\pm{\ifmmode\oldpm\else$\oldpm$\fi}
\newcommand\by{\ifmmode\times\else$\times$\fi}
\newcommand\ttt[1]{10\sp{#1}}
\newcommand\tttt[1]{\by\ttt{#1}}

\newcommand\tablebox[1]{\begin{tabular}[t]{@{}l@{}}#1\end{tabular}}
\newbox{\wdbox}
\renewcommand\c{\setbox\wdbox=\hbox{,}\hspace{\wd\wdbox}}
\renewcommand\i{\setbox\wdbox=\hbox{i}\hspace{\wd\wdbox}}
\newcommand\n{\hspace{0.5em}}

\catcode`@=11
\let\oldmsp=\sp
\let\oldmsb=\sb
\def\sp#1{\ifmmode
           \oldmsp{#1}%
         \else\strut\raise.85ex\hbox{\scriptsize #1}\fi}
\def\sb#1{\ifmmode
           \oldmsb{#1}%
         \else\strut\raise-.54ex\hbox{\scriptsize #1}\fi}
\newbox\@sp
\newbox\@sb
\catcode`@=12

\shorttitle{Characterization of WASP-8b with Spitzer}
\shortauthors{Cubillos {\em et al.}}

\slugcomment{\tablebox{Accepted for publication in {\em ApJ}. Submitted 2012 August 15; accepted 2013 March 5.}}

\begin{document}
\title{WASP-8\lowcase{b}: Characterization of a Cool and Eccentric Exoplanet with {\em Spitzer}}

\author{Patricio~Cubillos\altaffilmark{1,2},
        Joseph~Harrington\altaffilmark{1,2},
        Nikku~Madhusudhan\altaffilmark{3},
        Kevin~B.~Stevenson\altaffilmark{4},
        Ryan~A.~Hardy\altaffilmark{1},
        Jasmina~Blecic\altaffilmark{1},
        David~R.~Anderson\altaffilmark{5},
        Matthew~Hardin\altaffilmark{1}, and
        Christopher~J.~Campo\altaffilmark{1}}

\affil{\sp{1} Planetary Sciences Group, Department of Physics, 
       University of Central Florida, Orlando, FL 32816-2385}
\affil{\sp{2} Max-Plank-Institut f\"ur Astronomie, K\"onigstuhl 17,
  D-69117, Heidelberg, Germany}
\affil{\sp{3} Department of Physics and Department of Astronomy, Yale
  University, New Haven, CT 06511, USA}
\affil{\sp{4} Department of Astronomy and Astrophysics, University of Chicago, 5640 S. Ellis Ave, Chicago, IL 60637, USA}
\affil{\sp{5} Astrophysics Group, Keele University, Staffordshire ST5 5BG, UK}
\email{pcubillos@fulbrightmail.org}

\begin{abstract}
  WASP-8b has 2.18 times Jupiter's mass and is on an eccentric
  ($e=0.31$) 8.16-day orbit. With a time-averaged equilibrium
  temperature of 948 K, it is one of the least-irradiated hot Jupiters
  observed with the \SST.  We have analyzed six photometric light
  curves of WASP-8b during secondary eclipse observed in the 3.6, 4.5,
  and 8.0 \microns\ Infrared Array Camera bands.  The eclipse depths
  are $0.113\pm 0.018$\%, $0.069\pm 0.007$\%, and $0.093\pm 0.023$\%,
  respectively, giving respective brightness temperatures of 1552,
  1131, and 938 K.  We characterized the atmospheric thermal profile
  and composition of the planet using a line-by-line radiative
  transfer code and a Markov-chain Monte Carlo sampler. The data
  indicated no thermal inversion, independently of any assumption
  about chemical composition.  We noted an anomalously high
  3.6-\micron\ brightness temperature (1552~K); by modeling the
  eccentricity-caused thermal variation, we found that this
  temperature is plausible for radiative time scales less than
  $\sim\ttt2$ hours.  However, as no model spectra fit all three data
  points well, the temperature discrepancy remains as an open
  question.
\end{abstract}

\section{Introduction}
\label{introduction}

When transiting exoplanets pass behind their host stars (a secondary
eclipse), the observed flux drop provides a direct measurement of the
planet's thermal emission and reflected light. Today,
secondary-eclipse observations exist for nearly 30 exoplanets.  The
\SST\ \citep{WernerEtal2004apjsSpitzer} made most of these
observations, capturing broadband photometric light curves in six
near- and mid-infrared bands (3--24 \microns).  Each band probes a
specific altitude range in a planet's atmosphere.  With Bayesian
fitting of model spectra, one can quantitatively constrain the
atmospheric chemical composition and thermal profile of the planet's
photosphere \citep{MadhusudhanSeager2010}.  WASP-8b, with a
time-averaged equilibrium temperature of 948 \pm\ 22 K (\Teq,
temperature at which blackbody emission balances absorbed energy,
assuming zero albedo and efficient heat redistribution), is one of the
coolest Jupiter-sized planets yet observed in eclipse, and thus serves
as an end member to the set of measured hot-Jupiter atmospheres.

To classify the hot-Jupiter population, \citet{Fortney2008} proposed a
separation between moderately and strongly irradiated planets.  The
higher atmospheric temperatures of the more strongly irradiated
planets allow the presence of highly opaque molecules (like TiO and
VO) at high altitudes.  These strong absorbers produce hot
stratospheres (thermal inversion layers).  In contrast, for the
moderately irradiated hot Jupiters, these absorbers condense and rain
out to altitudes below the photosphere, thus presenting no thermal
inversions.

In general, the observations agree with this hypothesis, but
exceptions indicate that the picture is not yet completely understood.
For example, secondary-eclipse observations of the highly irradiated
WASP-12b \citep{Madhusudhan2011Nat, CrossfieldEtal2012apjWASP12b},
WASP-14b \citep{BlecicEtal2011}, and TrES-3
\citep{FressinEtal2010ApJ-Tres3} do not show evidence of thermal
inversions.  Conversely, XO-1 has an inversion layer even though it
receives a much lower stellar irradiation \citep{Machalek2008-XO-1b}.
Photochemistry provides one explanation.  The non-equilibrium
atmospheric chemistry models of \citet{ZahnleEtal2009SulfurPhotochem}
suggested that heating from sulfur compounds in the upper atmospheres
of hot Jupiters could explain these inversions.  Alternatively,
\citet{Knutson2010ApJ-CorrStarPlanet} suggest that strong UV radiation
from active stars destroys the high-altitude absorbers.

The Wide-Angle Search for Planets (WASP) Consortium discovered WASP-8b
in 2008 \citep{Queloz2010Wasp8}.  The planet orbits the brighter
component (WASP-8A) of a binary stellar system.  The angular
separation (4.83{\arcsec}) with the secondary (WASP-8B) sets a minimum
separation of 440 AU between the stars.  WASP-8A is a G6 star, with
effective temperature $T$\sb{eff} = 5600 K.  Color and photometric
analyses indicate that WASP-8B is a colder M star
\citep{Queloz2010Wasp8}.  WASP-8b is a 2.18 Jupiter-mass ($M$\sb{Jup})
planet with 1.08 times Jupiter's radius ($R$\sb{Jup}) in a retrograde
8.16 day orbit.  Its large eccentricity ($e = 0.31$) should make the
planet's dayside temperature vary by hundreds of degrees along the
orbit, possibly forcing an unusual climate.

The age of the host star (4 Gyr) is shorter than the planet's orbital
circularization time \citep[\sim30 Gyr, see, e.g.,][]{Goldreich1966Q,
  Bodenheimer2001}; accordingly, WASP-8b has one of the most eccentric
orbits among the \sim10-day-period exoplanets
\citep{PontEtal2011MNRASeccentricities}.  The Kozai mechanism
\citep{WuMurray2003Kozai} may explain the combination of high
eccentricity and retrograde orbit orientation.  The radial-velocity (RV)
drift and the large eccentricity may also indicate a second planetary
companion \citep{Queloz2010Wasp8}.

We obtained six secondary-eclipse light curves of WASP-8b from four
visits of the \SST, observing in the 3.6, 4.5, and 8.0 \microns\ bands
of the Infrared Array Camera \citep[IRAC, ][]{FazioEtal2004apjsIRAC}.
The eclipse depths determine the planet's dayside infrared emission.
Our Markov-chain Monte Carlo (MCMC)-driven radiative-transfer code
constrained the molecular abundances and temperature profile of
WASP-8b's dayside atmosphere, testing for the expected absence of a
thermal inversion and estimating the energy redistribution over its
surface.  We constrained the orbit of WASP-8b by determining the
eclipse epochs and durations.  We also modeled the thermal variations
along the orbit of the planet to explore the effects of the
eccentricity.

Section \ref{sec:observations} presents the \Spitzer\
observations of the WASP-8 system.
Section \ref{sec:analysis} describes the photometric and modeling
analysis of our secondary eclipse observations.
Section \ref{sec:orbit} gives the orbital dynamical analysis.
Section \ref{sec:atmosphere} presents our constraints on WASP-8b's
atmospheric composition derived from the photometry.
Section \ref{sec:discussion} discusses the effects of
eccentricity on the orbital thermal variation of WASP-8b.
Finally, Section \ref{sec:conclusions} states our conclusions.

\section{Observations}
\label{sec:observations}

The {\em Spitzer Space Telescope} visited WASP-8 four times.  From two
consecutive eclipse observations, we obtained simultaneous light
curves at 4.5 and 8.0 \microns.  Later, from two more consecutive
eclipse observations during the {\em Warm} \Spitzer\ mission, we
obtained one light curve at 3.6 \microns\ and one at 4.5 \microns\
(see Table \ref{table:observations}).  The \Spitzer\ pipeline (version
18.18.0) processed the raw data, producing Basic Calibrated Data
(BCD).

\begin{figure}[tb]
  \centering
  \includegraphics[width=\linewidth, clip, trim =70 30 100 20]{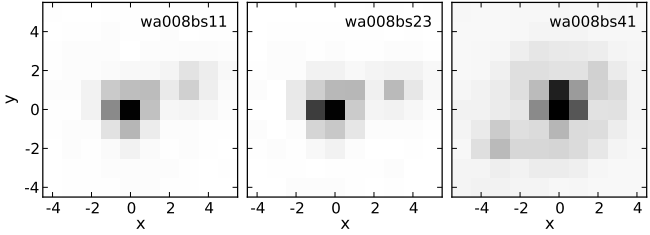}
  \caption{\Spitzer\ images of WASP-8 at 3.6, 4.5, and 8.0 \microns,
    respectively. The brighter star, WASP-8A, is at the origin. The
    dimmer WASP-8B signal overlapped that of WASP-8A.}
  \label{fig:SpitzerFOV}
\end{figure}

\begin{table}[ht]
\centering
\caption{\label{table:observations} Observation Information}
\strut\hfill
\setlength{\tabcolsep}{3pt}
\begin{tabular}{cccccc}
    \hline
    \hline
    Label\tablenotemark{a}
              & Wavel.   & Observation & Duration  & Exp. time & Cadence \\ 
              & \microns & date        &  minutes  & seconds   & seconds \\ 
    \hline               
    \waspt2   & 4.5      &  2008-12-13 & 226       &   \n1.20  & \n2.0   \\ 
    \waspf2   & 8.0      &  2008-12-13 & 226       &    10.40  &  12.0   \\ 
    \waspt1   & 4.5      &  2008-12-21 & 226       &   \n1.20  & \n2.0   \\ 
    \waspf1   & 8.0      &  2008-12-21 & 226       &    10.40  &  12.0   \\ 
    \waspo    & 3.6      &  2010-07-23 & 458       &   \n0.36  & \n0.4   \\ 
    \waspt3   & 4.5      &  2010-07-31 & 458       &   \n0.36  & \n0.4   \\ 
    \hline
\end{tabular}
\hfill\strut
\tablenotetext{1}{
{\it wa008b} designates the planet, {\it s} specifies secondary
eclipse, and the two numbers indicate the wavelength channel and
observation serial number (we analyzed the 2008-12-21 data before the
2008-12-13 data and inadvertently inverted the serial numbers).}
\end{table}

During the initial minutes of our observations, the telescope pointing
drifted \sim0.25 pixels before stabilizing.  Throughout the
observations, the pointing also jittered from frame to frame
($\sim0.01$ pixel) and oscillated in an hour-long periodic movement
($\sim0.1$ pixel amplitude).

The separation between the centers of WASP-8A and WASP-8B in the IRAC
detectors is only 3.7 pixels. Consequently, the signal from the stars
overlapped, demanding special care during the data analysis (see
Figure \ref{fig:SpitzerFOV}).  Table \ref{table:WASPsystem} shows the
average and standard deviation of the flux ratio, separation, and
position angle (PA) of the secondary star with respect to WASP-8A
(derived from our centering routine, see Section
\ref{ss:center}). Our PA values agree with those of
\citet{Queloz2010Wasp8}, but our separation values are consistently
lower than theirs ($4.83{\arcsec} \pm 0.01${\arcsec}).

\begin{table*}[ht]
\centering
\caption{\label{table:WASPsystem} WASP-8 System}
\begin{tabular}{cr@{\hspace{0.25cm}}l@{\hspace{0.75cm}}r@{\hspace{0.25cm}}l@{\hspace{0.75cm}}r@{\hspace{0.25cm}}l@{\hspace{0.75cm}}r@{\hspace{0.25cm}}l}
    \hline
    \hline
Event    & \multicolumn{2}{l}{\phantom{de}Flux ratio} & \multicolumn{2}{l}{Separation (pixels)} & \multicolumn{2}{l}{Separation (\arcsec)} & \multicolumn{2}{l}{Position angle (deg)}  \\
         & average & stddev  & average & stddev           & average & stddev           & average & stddev                 \\
    \hline
\waspo   & 0.1420 & 0.0030  & 3.760 & 0.013   & 4.610 & 0.016   & 171.32 & 0.28    \\
\waspt1  & 0.1512 & 0.0017  & 3.734 & 0.007   & 4.541 & 0.009   & 171.00 & 0.15    \\
\waspt2  & 0.1600 & 0.0022  & 3.737 & 0.009   & 4.544 & 0.011   & 170.75 & 0.16    \\
\waspt3  & 0.1648 & 0.0039  & 3.726 & 0.017   & 4.497 & 0.020   & 170.78 & 0.29    \\
\waspf1  & 0.1718 & 0.0020  & 3.690 & 0.009   & 4.513 & 0.011   & 170.84 & 0.17    \\
\waspf2  & 0.1794 & 0.0023  & 3.686 & 0.010   & 4.506 & 0.012   & 170.96 & 0.16    \\
    \hline
\end{tabular}
\end{table*}

\section{Data Analysis}
\label{sec:analysis}

Our Photometry for Orbits, Eclipses, and Transits (POET) pipeline
produces light curves from BCD images.  Briefly, POET creates a bad
pixel mask for each image, finds the center position of the target,
executes interpolated aperture photometry, and fits a light curve
model that includes physical and systematic parameters.

\subsection{POET: Initial Reduction}

POET created bad pixel masks by discarding the flagged pixels from the
{\Spitzer} BCD masks. Then, it discarded outlier pixels with a
sigma-rejection method.  At each pixel position and in sets of 64
consecutive images, POET calculated the median and standard deviation
of the unmasked pixels. Pixels diverging more than four times the
standard deviation from the median were masked. We iterated this
process twice.

We obtained the Julian Date of each frame from the UTCS\_OBS and
FRAMTIME entries of the files' headers.  We calculated the Barycentric
Julian Date (BJD) by correcting the projected light-travel time from
the telescope to the Solar System's barycenter using the Jet
Propulsion Laboratory (JPL) Horizons system.  We report the times in
both Coordinated Universal Time (UTC) and the Barycentric Dynamical
Time (TDB); the latter is unaffected by leap seconds
\citep{Eastman2010}.

\subsection{Centering}
\label{ss:center}

POET provides three routines to determine the center of the
point-spread function (PSF) in each image: center of light,
2D-Gaussian fitting, and least asymmetry \citep[Supplementary
Information]{StevensonEtal2010Natur}.  The proximity of WASP-8B
confuses these methods, so we added a double-PSF fit that shifts
supersampled PSFs to the target and secondary, bins them down, and
scales their amplitudes, as in \citet{Crossfield2010}.  For each
\Spitzer\ band we used Tiny
Tim\footnote{http://irsa.ipac.caltech.edu/data/SPITZER/docs/dataanaly-sistools/tools/contributed/general/stinytim/} (version 2.0) to create a stellar PSF
model with a 5600 K blackbody spectrum at 100{\by} finer resolution
than our images.  The double-PSF routine has seven free parameters:
the position of each star ($x\sb A$, $y\sb A$, $x\sb B$, $y\sb B$),
the integrated stellar fluxes ($F\sb A$, $F\sb B$), and the background
sky flux ($f$\sb{sky}).

To avoid interpolation when binning down, the PSF shifts are quantized
at the model's resolution, such that image and model pixel boundaries
coincide.  This quantization sets the position precision to 0.01
pixels.  It also excludes the position parameters from \chisq\
minimizers that assume a continuous function, such as
Levenberg-Marquardt.  So, we fit $F\sb A$, $F\sb B$, and $f$\sb{sky}
for a given position set \textbf{\emph{x}}~$ = \{x\sb A$, $y\sb A$,
$x\sb B$, $y\sb B\}$.

To avoid the computational challenge of performing a \chisq\
minimization for each \textbf{\emph{x}} in a 4D space at 0.01 pixel
resolution, we explored only specific coordinate positions.  Starting
at an initial guess position, and with an initial jump step of 100
positions (1 image pixel), we calculated \chisq\ at that position and
the 80 ($=3\sp4-1$) adjacent positions that are one jump step away
along all combinations of coordinate directions.  We either moved to
the lowest \chisq or, if already there, shrank the step by half.  We
repeated the procedure until the step was zero.

\subsection{Photometry}

Circular aperture photometry is unsuitable for this system, since any
flux from the secondary star (WASP-8B) contained in the aperture
dilutes the eclipse depth of WASP-8b.  Small pointing jitter would
also increase the light-curve dispersion for any aperture that
included much WASP-8B flux.  Apertures that are too large or small
both produce noisier light curves.  Thus, we modified the POET
interpolated aperture photometry \citep[Supplementary
Information]{HarringtonEtal2007natHD149026b} to remove the secondary
star two different ways.  In both methods, we subtracted the median
sky level prior to the stellar flux calculation.  The sky annulus
included values 7 -- 15 pixels from the target.

In our first method (B-Subtract), we subtracted the fitted, binned PSF
model of WASP-8B from each image. Then, we performed interpolated
aperture photometry centered on the target (A aperture).  In the
second method (B-Mask), we discarded the pixels within a circular
aperture centered at the position of the secondary before performing
aperture photometry.  The mask's aperture must encompass most of the
contribution from WASP-8B, but not from the target. Therefore, we
tested mask apertures with 1.6, 1.8, and 2.0 pixel radii.  For each
photometry method we tested a broad range of A-aperture radii in 0.25
pixel intervals.

The B-Mask method has less residual dispersion when the mask is
located at a fixed vector separation from WASP-8A (using the median of
all the measured separations in an event), than when its position is
determined for each individual frame.  This can be explained by the
dimmer signal of WASP-8B, which lowers the accuracy of its position
estimation.  So, within each data set using B-Mask, we used the median
vector separation of the two objects.  For the B-subtract method, the
standard deviation of the normalized residuals (SDNR) and
eclipse-depths differences are marginal.

\subsection{Light Curve Modeling}

The eclipse depths of WASP-8b are on the order of 0.1\% of the
system's flux, well below {\em Spitzer's}\/ photometric stability
criteria \citep{FazioEtal2004apjsIRAC}. Thus, the eclipse light-curve
modeling requires a thorough characterization of the detector
systematics.  Systematic effects have been largely observed and
documented; they can have both temporal and spatial components, and
vary in strength and behavior for each data set.

The main systematic at 3.6 and 4.5 {\microns} is intrapixel
sensitivity variation, $M(x,y)$, where the measured flux depends on
the precise position of the target on the array
\citep{StevensonEtal2012apjHD149026b, CharbonneauEtal2005apjTrES1}.
In addition, the detectors show a time-dependent sensitivity variation
called the \emph{ramp} effect, $R(t)$, suspected to be caused by
charge trapping \citep{Agol2010ApjHD189} at 8.0 \microns, but there
are also reports of a ramp in the 3.6 and 4.5 \micron\ bands
\citep[e.g., ][]{Campo2011, NymeyerEtal2011, Knutson2011gj436,
  Deming2011Corot, BlecicEtal2011, StevensonEtal2010Natur,
  StevensonEtal2012apjHD149026b}.  The eclipse and both systematic
variations entangle to produce the observed light curve.  To account
for their contributions, we modeled the light curves as
\begin{equation}
F(x,y,t) = F\sb s M(x,y) R(t) E(t),
\label{eq:fire}
\end{equation}
where $F\sb s$ is the out-of-eclipse system flux.  We used the eclipse
model, $E(t)$, from \citet{MandelAgol2002apjLightcurves}.  The eclipse
is parametrized by the eclipse depth, the mid-point phase, the
duration, and the ingress and egress times.  For the ingress/egress
times we adopted a value of 18.8 min, derived from the orbital
parameters of the planet. We used this value in all of our
eclipse-model fits.

The strength and behavior of the ramp variations are specific to each
dataset. Many formulae have been applied in the literature
\citep[e.g.,][]{DemingEtal2007, HarringtonEtal2007natHD149026b,
  Knutson2011gj436, StevensonEtal2012apjHD149026b}. The models are
formed with combinations of exponential, logarithmic, and polynomial
functions.  We tested dozens of equations; the best were:
\begin{eqnarray}
\label{eqnre}
{\rm risingexp}:     \quad R(t) & = & 1 - e\sp{-r\sb{0}(t-t\sb{0})} \\
\label{eqnlog}
{\rm logramp}\sb q:  \quad R(t) & = & 1 + r\sb{q} [\ln(t-t\sb{0})]\sp q \\
\label{eqnlin}
{\rm linramp}:      \quad R(t) & = & 1 + r\sb{1}(t-t\sb c) \\
\label{eqnquad}
{\rm quadramp}\sb 1: \quad R(t) & = & 1 + r\sb{1}(t-t\sb c) + r\sb{2}(t-t\sb c)\sp{2} \\
\label{eqnquad2}
{\rm quadramp}\sb 2: \quad R(t) & = & 1 + r\sb{2}(t-t\sb c)\sp{2} \\
\label{eqnll}
{\rm loglinear}:     \quad R(t) & = & 1 + r\sb{1}(t-t\sb c) + r\sb{4} \ln(t-t\sb{0})
\end{eqnarray}
where $t\sb c$ is a constant value at the approximated mid-point
phase of the eclipse ($t\sb c = 0.515$ for this planet). Slight
changes in $t\sb c$ do not significantly affect the fitted eclipse
parameters.

We used our Bi-Linearly Interpolated Subpixel Sensitivity (BLISS)
mapping technique \citep{StevensonEtal2012apjHD149026b} to calculate
$M(x,y)$.  The BLISS method has been found to return a better result
than a polynomial fit \citep{StevensonEtal2012apjHD149026b,
  BlecicEtal2011}.

To determine the best-fitting parameters of our model, Equation
(\ref{eq:fire}), we used a $\chi\sp2$ minimizer with the
Levenberg-Marquardt algorithm.  We used Bayesian posterior sampling
via an MCMC algorithm to explore the phase
space and estimate the uncertainties of the free parameters of the
light-curve models.  Our code implements the Metropolis random walk,
which proposes parameter sets from a multivariate normal distribution
centered at the current position in the chain, computes
$\chi\sp{2}$, and accepts (or rejects) the new set with greater
probability for a lower (higher) $\chi\sp{2}$.  By generating
millions of parameters sets, the algorithm samples the posterior
distribution of the model parameters.  As a necessary condition for
chain convergence, we require the Gelman-Rubin statistic
\citep{GelmanRubin1992} to be within 1\% of unity for each free parameter
between four MCMC chains.

The photometry routine uses the BCD uncertainty images to estimate the
uncertainties of the light-curve data points, $\sigma\sb i$.  However,
since the \Spitzer\ pipeline in general overestimates these
uncertainties (it is designed for absolute photometry), we multiply by
a constant factor ($\sigma\sb i \rightarrow f \cdot \sigma \sb{i}$),
such that the reduced $\chi\sp2 = 1$ in the light-curve fit.  This is
equivalent to estimating a single $\sigma$ from the scatter of model
residuals.  Both methods account for red noise, but ours retains the
(usually small) $\sigma\sb i$ variations due to aberrant frames.

To determine the best raw light curve (i.e., by selection of
photometry method and aperture radius), we calculated the SDNR of the
light-curve fit \citep{StevensonEtal2012apjHD149026b, Campo2011}. Poor
fits or data with high dispersion increase SDNR; the optimum data set
minimizes the SDNR value.  Once we chose the best light curve, we
compared the different ramp models according to the Bayesian
Information Criterion \citep{Liddle2007},
\begin{equation}
{\rm BIC} = \chi\sp2 + k\ln N,
\end{equation}
where $k$ is the number of free parameters and $N$ the number of data
points.  The best model minimizes the BIC.  The probability ratio
favoring one model over a second one is $\exp(-\Delta{\rm BIC}/2)$.

\subsubsection{wa008bs11 Analysis} 
\label{sss:wa11}

This observation started 2.9 hours before the eclipse's first contact.
The telescope observed the target in sub-array mode, allowing a high
cadence (Table \ref{table:observations}).  We discarded the initial 15
minutes of observation while the telescope pointing settled.  Our data
present both intrapixel and weak ramp systematics.

\begin{figure*}[tb]
\strut\hfill 
\includegraphics[width=0.315\textwidth,trim=20 0 21 48,clip]
{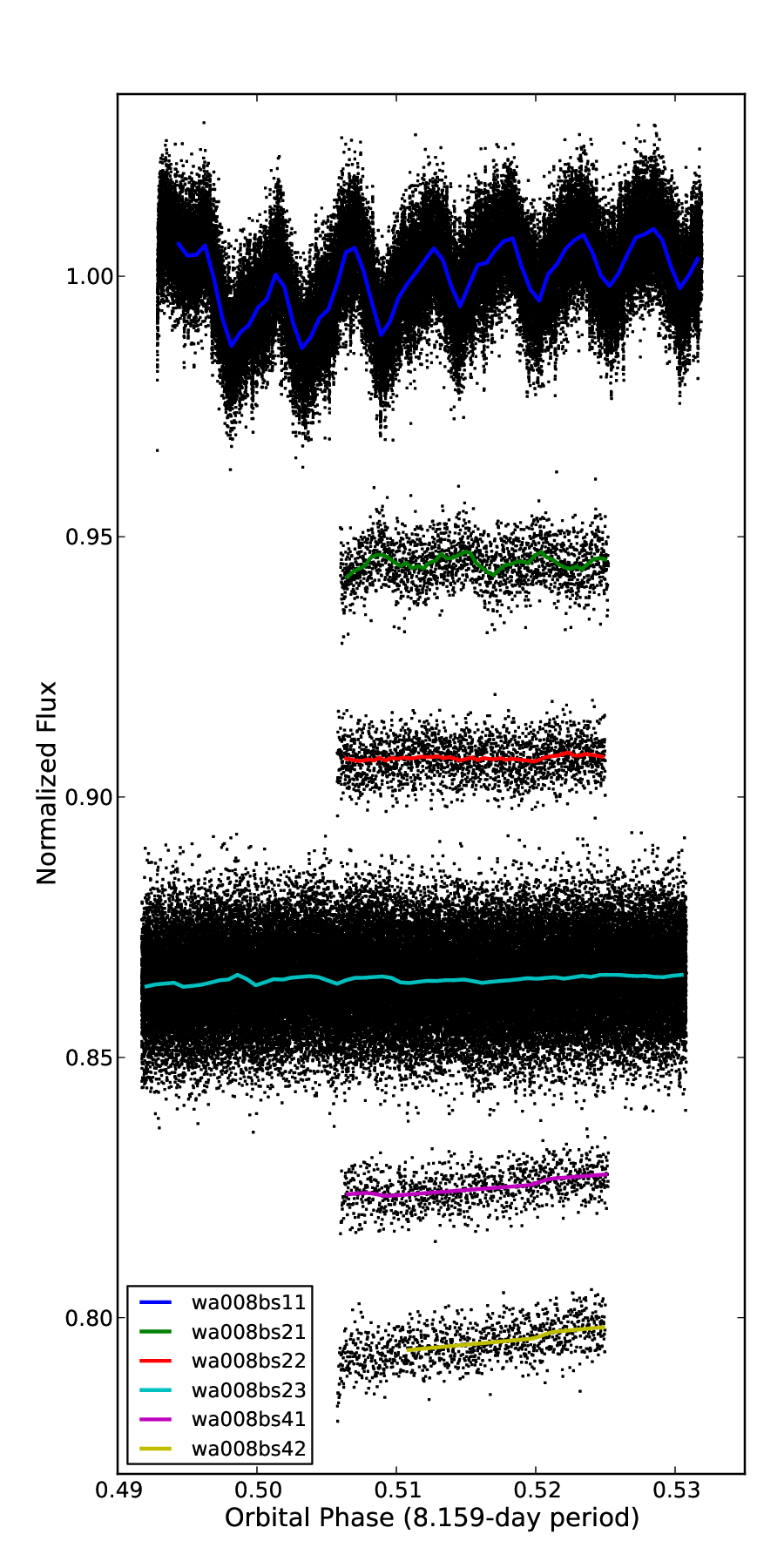}\hfill
\includegraphics[width=0.315\textwidth,trim=20 0 21 48,clip]
{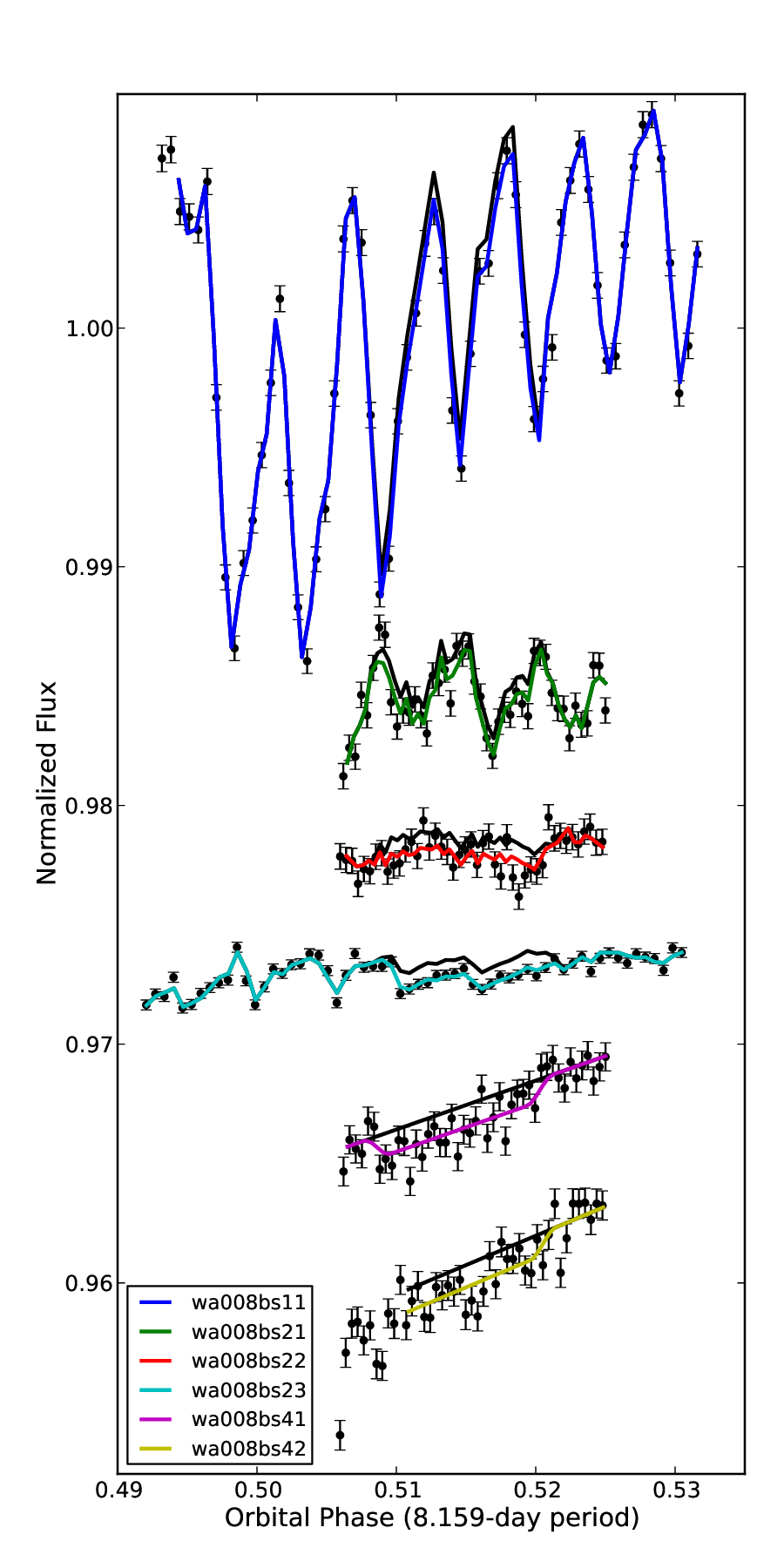}\hfill
\includegraphics[width=0.32\textwidth,trim=12 0 21 48,clip]
{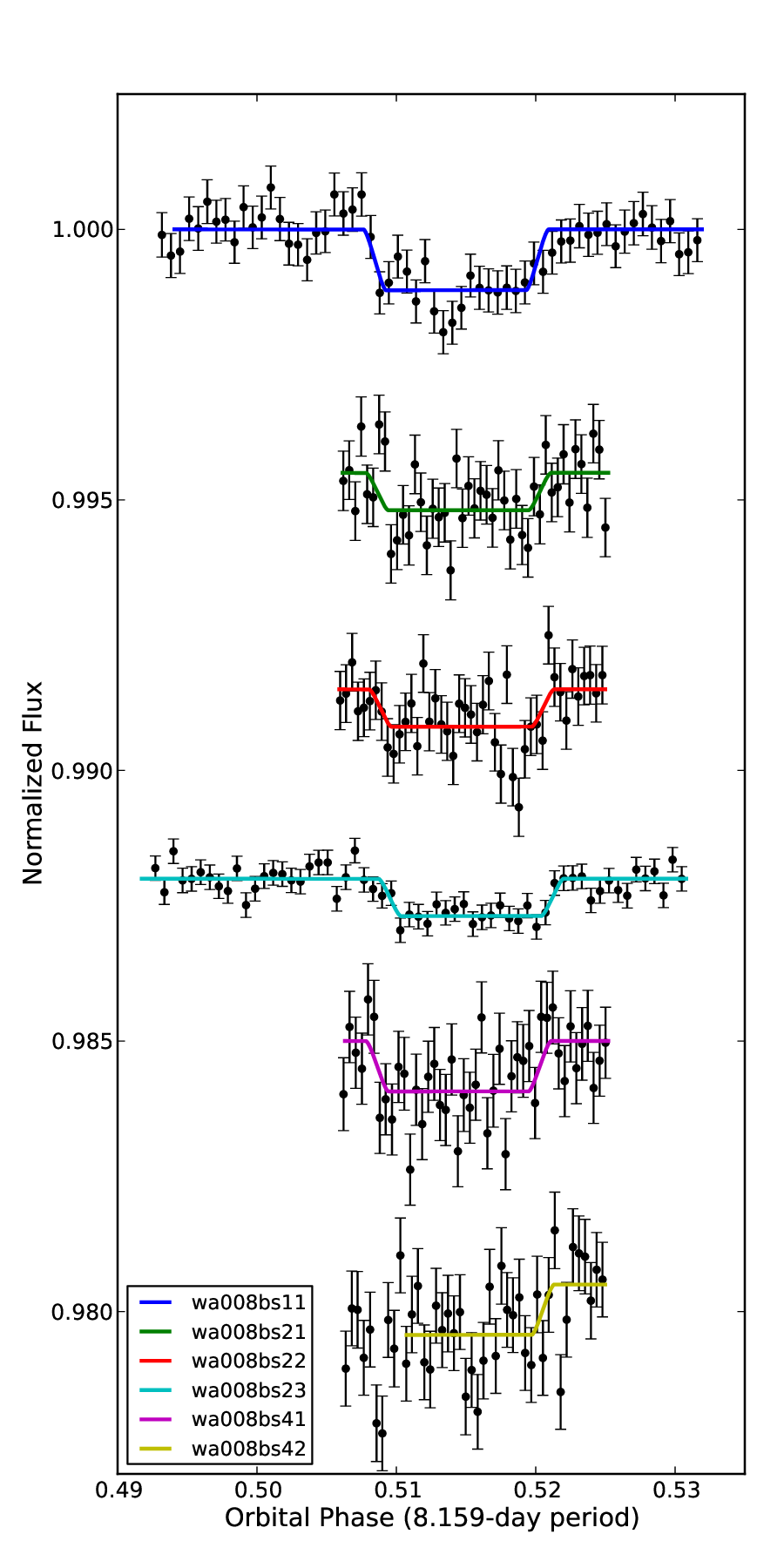}\hfill\strut
\caption{\label{fig:lightcurves} Raw (left), binned (center) and
  systematics-corrected (right) secondary-eclipse light curves of
  WASP-8b at 3.6, 4.5, and 8.0 {\microns}.  The system flux is
  normalized to unity and the points are shifted
  vertically for clarity.  The colored
  curves are the best-fit models (see legend).  The black curves are
  the best-fit models excluding the eclipse component. The error bars
  in the center and right panels are the 1$\sigma$
  uncertainties.}
\end{figure*}

\begin{table}[ht]
\caption{\label{table:wa008bs11ramps} 
\waspo\ Ramp Model Fits}
\strut\hfill\begin{tabular}{cccccc}
    \hline
    \hline
    $R(t)$        &  SDNR      & $\Delta$BIC & Ecl. Depth (\%) \\
    \hline
    quadramp\sb1  &  0.0061141 & \n0.00      & 0.119 \\
    risingexp     &  0.0061148 & \n1.73      & 0.106 \\
    logramp\sb1   &  0.0061153 & \n2.96      & 0.096 \\
    linramp       &  0.0061201 & \n6.34      & 0.063 \\
    loglinear     &  0.0061141 &  11.10      & 0.119 \\
    \hline
\end{tabular}\hfill\strut
\end{table}

The 2.25 pixel A aperture with B-subtract photometry minimized SDNR.
Table \ref{table:wa008bs11ramps} shows the five best-fitting models to
the best \waspo\ light curve. $\Delta$BIC is with respect to the
lowest BIC value.  The quadramp\sb{1} model is 2.4 times more probable
than, and consistent with, the second-best model. The linear
($11\sigma$) and and quadratic ($4\sigma$) terms of the quadramp\sb{1}
model (see Table \ref{table:fits}) confirm the need for a ramp model.
As a general remark, we noted that all the logramp\sb{q} models
produce similar BIC and eclipse parameter values; therefore, we will
refer only to the logramp\sb{1} model in the future. Models with more
free parameters do not improve BIC.  Following
\citet{StevensonEtal2012apjHD149026b}, we vary the bin size and the
minimum number of data points per bin (\mnp) of the BLISS map to
minimize the dispersion of the residuals. We required at least 4
points per bin for any dataset.  The PSF-fitting position precision of
0.01 pixels sets our lower limit for the binsize.  For \waspo, $mnp =
5$ and a bin size of 0.015 pixels optimized the fit.

\begin{figure}[tb]
\centering
\includegraphics[width=0.95\linewidth, clip, trim=0 40 5 30]{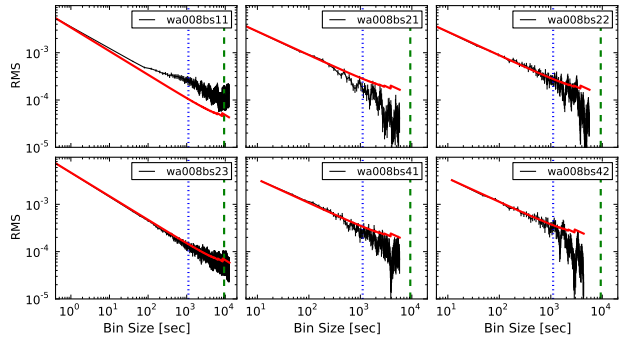}
\caption{RMS of the fit residuals (black curves with 1$\sigma$
  uncertainties) {\vs} bin size of the WASP-8b light curves. The red
  curves are the expected RMS for Gaussian noise (extrapolated
  unbinned RMS scaling by the inverse square root of the bin
  size). The blue dotted and green dashed vertical lines mark the bin
  size corresponding to the eclipse ingress and duration time,
  respectively. \waspo's excess above the red line indicate correlated
  noise at bin sizes larger than the ingress time.}
\label{fig:rms}
\end{figure}

Figure \ref{fig:lightcurves} shows the raw, binned and
systematics-corrected \waspo\ light curves with their best-fitting
model.  We considered the correlated noise in the residuals as well
\citep{Pont2006Rednoise}.  Figure \ref{fig:rms} shows the
root-mean-square (RMS) of the residuals \vs bin size. The \waspo\ RMS
curve deviates above the expected RMS for pure Gaussian noise.
Following \citet{WinnEtal2008Rednoise}, to account for the correlated
noise, we weighted the light curve uncertainties by the factor $\beta$
(the fractional RMS excess above the pure Gaussian RMS at the bin size
corresponding to the eclipse duration). For \waspo\ we found $\beta
=2.4$.  We inspected all the pairwise correlation plots and histograms
and found only unimodal Gaussian distributions.

Alternatively, the residual-permutation (also known as {\em prayer
  bead}) algorithm is sometimes used to assess correlated noise in a
fit.  In this method, we cyclically shift the residuals from the best
model by one frame, add them back to that model, and re-fit, repeating
until we shift the residuals back to their original positions.  This
generates a distribution of values for each parameter, from which we
estimate the parameter uncertainties.  The eclipse-depth uncertainty
is 0.021\%, similar to the value found with the
\citet{WinnEtal2008Rednoise} method (see Table \ref{table:fits}).
However, we are cautious.  Although {\em prayer bead} has been broadly used
for the analysis of exoplanet lightcurve and RV fits
\citep[e.g.,][]{Southworth2008HomogTransit, BouchyEtal2005CorrNoise,
  PontEtal2005, GillonEtal2007Gj436, Knutson2008HD209,
  CowanEtal2012WASP12b}, we have found no detailed description of its
statistical properties in the literature.

\subsubsection{wa008bs23 Analysis}
\label{sss:wa23}

\begin{figure}[htb]
\centering
\includegraphics[width=\linewidth, clip, trim=5 30 35 105]{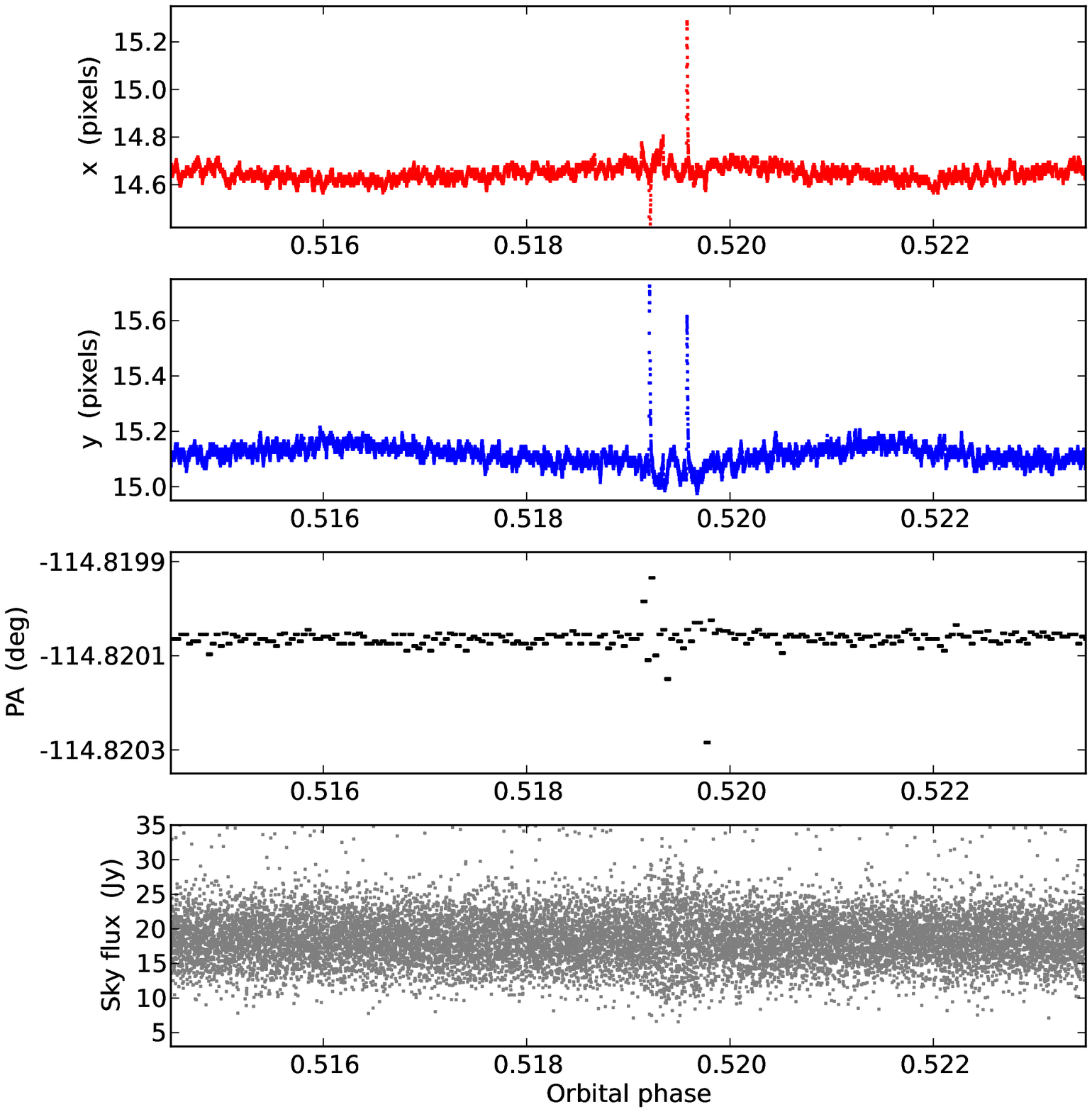}
\caption{\waspt3 target pointing, position angle and background sky
  flux near phase 0.519. We observed two sudden position shifts
  coincident with increases in the background flux. All values
  returned to normal almost instantly.}
\label{fig:impact}
\end{figure}

With the same observing setup as \waspo, this observation started 3.3
hours prior to the eclipse's first contact.  This dataset also
presented both intrapixel and ramp variations. Note that the
intrapixel systematic is weaker than at 3.6 \microns, attributed to
the smaller degree of undersampling at larger wavelengths by
\citet{CharbonneauEtal2005apjTrES1} and
\citet{Morales-CalderonEtal2006apjIntrapixel}. Even though the
pointing stabilized only after the initial 20 minutes, the light curve
did not deviate significantly; therefore, we included all data points
in the analysis. We noted two sudden pointing and PA deviations near
phase 0.519. After each incident, the telescope resumed its position
within \sim10 s (Figure \ref{fig:impact}).  Micrometeorite
impacts on the telescope can explain the abrupt deviations.
Simultaneously, we measured a slight increase in the background sky
flux dispersion, which returned to normality shortly after.  The
target flux did not show any extraordinary fluctuations during these
incidents.  However, the points outside the normal pointing range were
eliminated by the BLISS map's {\mnp} criterion.

The analysis is analogous to \waspo. The SDNR indicated clearly that
the 2.25 pixel A aperture with B-Subtract photometry produced the
lowest dispersion.  The best-fitting ramp model is logramp\sb{1},
which is 21 times more probable than the rising exponential ramp
(Table \ref{table:wa008bs23ramps}).  The BLISS map is optimized at
$mnp=4$ and a bin size of 0.025 pixels.

An initial MCMC run showed a significant linear correlation between
the system flux and the $r\sb{1}$ parameter of the logramp, which
prevented the MCMC chain from converging.  We solved this problem by
transforming the correlated parameters into an orthogonal set of
parameters, rerunning the MCMC chain, and inverting the transformation
on the resulting parameter values
\citep{StevensonEtal2012apjHD149026b}.  Figures \ref{fig:lightcurves}
and \ref{fig:rms} show the \waspt3 light curves with the best-fitting
model and RMS of the residuals \vs bin size.

\begin{table}[ht]
\caption{\label{table:wa008bs23ramps} 
\waspt3 Ramp Model Fits}
\strut\hfill\begin{tabular}{cccccc}
    \hline
    \hline
    $R(t)$        &  SDNR        & $\Delta$BIC & Ecl. Depth (\%) \\
    \hline
    logramp\sb1   &  0.0073830   & \n0.00      & 0.0677   \\
    risingexp     &  0.0073832   & \n6.10      & 0.0730   \\
    quadramp\sb1  &  0.0073833   & \n9.40      & 0.0777   \\
    loglinear     &  0.0073830   &  10.84      & 0.0685   \\
    linramp       &  0.0073846   &  12.19      & 0.0564   \\
    \hline
\end{tabular}\hfill\strut
\end{table}

\subsubsection{wa008bs21 \& wa008bs41 Analysis}

We simultaneously observed \waspt1 and \waspf1 in full-array mode.
Prior to the eclipse observation, we exposed the detector (a
``preflash'' observation, \citealp{KnutsonEtal2009apjHD149026bphase})
for 25 minutes to a bright H\,{\footnotesize II} region, with coordinates
$\alpha~=~20$\sp{h} 21\sp{m} 39\sp{s}.28 and
$\delta~=~+37$\sp{\degrees} 31{\arcmin} 03.6{\arcsec}, to minimize the
ramp systematic variation.  The secondary-eclipse observation started
only 26 minutes before the first contact. The telescope pointing
stabilized quickly, so fortunately we needed to remove only the
initial four minutes of observation.  Every 12 s, the detector
recorded two consecutive images (two-second exposures) at 4.5
{\microns} and one image at 8.0 {\microns}. (Table
\ref{table:observations}).

\begin{figure}[ht]
\centering
\includegraphics[width=\linewidth, clip, trim=0 0 50 30]{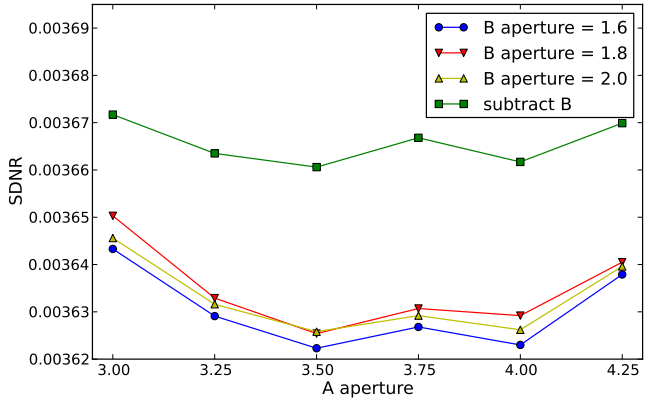}
\caption{\waspt1 Standard Deviation of the Normalized Residuals
  \emph{vs.}\ aperture (in pixels). The SDNR curves use the best ramp
  model in Table \ref{table:wa008bs21ramps}. The legend indicates the
  photometry method.  SDNR increases at 3.75 pixels (coincident with
  the stars' separation).  The eclipse parameters are consistent over
  the 3.0--4.25 aperture range.  The optimum dataset uses 1.6 pixel
  B-Mask photometry with a 3.5 pixel A aperture.}
\label{fig:sdnr21}
\end{figure}

The SDNR analysis of \waspt1 showed that a 3.5 pixel A aperture with
1.6 pixel B-Mask photometry minimizes the dispersion (Figure
\ref{fig:sdnr21}).  The ramp models indicated a negligible ramp
variation.  Accordingly, a fit without a ramp model yielded the lowest
BIC.  Table \ref{table:wa008bs21ramps} shows the four best-fitting
models for the best \waspt1 data set.  The no-ramp model is 15 times
more probable than the quadramp\sb2 model.

\begin{table}[ht]
\caption{\label{table:wa008bs21ramps} 
\waspt1 Ramp Model Fits}
\strut\hfill\begin{tabular}{cccccc}
    \hline
    \hline
    $R(t)$          & SDNR      & $\Delta$BIC & Ecl. Depth (\%) \\
    \hline
    no ramp         & 0.0036223 & \n0.00      & 0.0718   \\
    quadramp\sb2    & 0.0036195 & \n5.76      & 0.1189   \\
    linramp         & 0.0036223 & \n7.56      & 0.0714   \\
    quadramp\sb1    & 0.0036197 &  13.14      & 0.1170   \\
    \hline
\end{tabular}\hfill\strut
\end{table}

\begin{figure}[htb]
\centering
\includegraphics[width=\linewidth, clip]{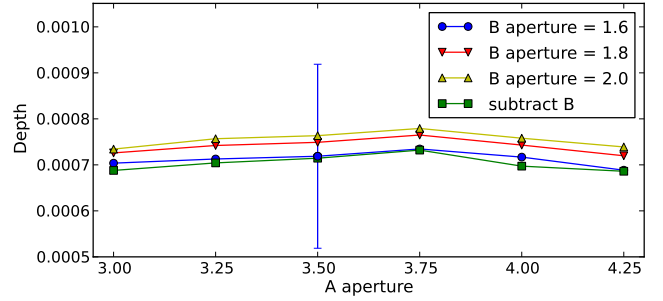}
\caption{Eclipse depth \emph{vs.} A aperture for \waspt1. Each color
  represent a different photometry method as in Figure
  \ref{fig:sdnr21}.  The blue error bar corresponds to the 1-$\sigma$
  uncertainty of the best model.  The eclipse duration and mid-point
  phase follow a similar trend.}
\label{fig:depth21}
\end{figure}

Because of the shorter out-of-eclipse observation, the system flux is
less-constrained for \waspt1\ than for the \waspo\ or \waspt3 events.
Combined with a correlation between the eclipse depth and system flux
(revealed by MCMC), the lower precision of the system flux translates
into a larger eclipse depth uncertainty.  Nevertheless, the \waspt1
fit parameters were consistent among the different apertures (Figure
\ref{fig:depth21}). The optimum parameters of the 
BLISS map are $mnp=5$ and a bin size of 0.025 pixels.

\begin{figure}[tb]
\centering
\includegraphics[width=\linewidth, clip, trim=0 0 50 30]{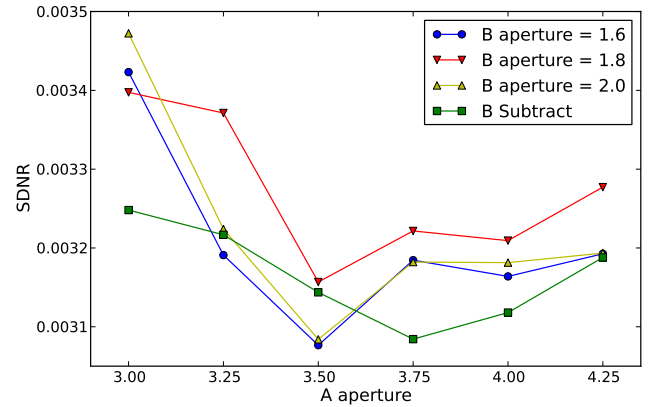}
\caption{Joint \waspt1+\waspf1 Standard Deviation of the Normalized
  Residuals \emph{vs.}\ aperture radius (in pixels) of \waspf1, for
  different photometry methods.  All 24 light curves use the best ramp
  model from Table \ref{table:wa008bs41ramps}.  Light curves using 1.6
  and 2.0 pixel B-Mask photometry at 3.5 pixel A aperture produce
  consistent eclipse parameters and outperform the best B-subtract
  method (with a 3.75 pixel A aperture). The best B-subtract also
  yields a more scattered raw light curve. Hence, the optimum dataset
  uses 1.6 pixel B-mask photometry with a 3.5 pixel A aperture.}
\label{fig:sdnr41}
\end{figure}

The 8.0 \microns\ detector did not present an intrapixel pattern like
the 3.6 or 4.5 \microns\ detectors.  However, some of the raw light
curves for different apertures and photometry methods showed large
scatter and presented strong oscillations, producing implausible fit
parameters.  A pixelation effect \citep{Anderson2011MNRASWasp17b,
  StevensonEtal2012apjHD149026b} might be responsible.  As a
consequence, we were unable to fit the eclipse parameters
unambiguously for this data set alone.  Normally we study the events
individually to select the best aperture and photometry method, but in
this case we used a joint fit with the best \waspt1 dataset and model
to help constrain the 8.0 {\micron} eclipse curve, sharing the eclipse
duration and mid-point parameters.  The 3.5 pixel A aperture with 1.6
pixel B-mask photometry for \waspf1 minimized the joint SDNR (Figure
\ref{fig:sdnr41}).

Table \ref{table:wa008bs41ramps} compares the four best-fitting ramp
models for the best \waspf1 light curve. A linear ramp minimizes BIC,
and is 20 times more probable than the next-best model.  Figures
\ref{fig:lightcurves} and \ref{fig:rms} show the \waspt1 and \waspf1
light curves with their best-fitting models and RMS of the residuals
\vs bin size, respectively.

\begin{table}[ht]
\caption{\label{table:wa008bs41ramps} 
\waspf1 Ramp Model Fits}
\strut\hfill\begin{tabular}{ccccc}
    \hline
    \hline
    R(t)          & SDNR 4\&2 & $\Delta$BIC & Ecl. Depth (\%)\\
    \hline
    linramp       & 0.0030766 &  0.00            & 0.0931   \\
    quadramp\sb1  & 0.0030744 &  5.94            & 0.1308   \\
    risingexp     & 0.0030754 &  6.33            & 0.1150   \\
    logramp\sb1   & 0.0030758 &  6.50            & 0.1078   \\
    \hline
\end{tabular}\hfill\strut
\end{table}

\subsubsection{wa008bs22 \& wa008bs42 Analysis}

The observing setup of these events was identical to \waspt1 and
\waspf1, including the preflash observation.  The pointing of this
observation drifted noticeably more than in the other observations,
moving more than $0.4$ pixels during the initial 30 minutes and
stabilizing only during the eclipse.  As a consequence, the
illumination level of the individual pixels changed during the
beginning of the eclipse. The ramp variation, which depends on the
illumination \citep{Knutson2008HD209}, was disrupted.

The \waspt2 event, having a negligible ramp variation, was little
affected by the telescope pointing shift.  The SDNR calculation for
\waspt2 indicated the 1.8 pixel B-Mask photometry with 3.75 pixel A
aperture as the best dataset.  A light-curve model without a ramp
(Table \ref{table:wa008bs22ramps}) is 639 times more probable than the
quadramp\sb2 model. The optimal BLISS map has $mnp=4$ and a bin size
of 0.02 pixels.

\begin{table}[ht]
\caption{\label{table:wa008bs22ramps} 
\waspt2 Ramp Model Fits}
\strut\hfill\begin{tabular}{cccccc}
    \hline
    \hline
    $R(t)$        & SDNR      & $\Delta$BIC & Ecl. Depth (\%) \\
    \hline
    no ramp       & 0.0025274 &  \n0.00     & 0.0814  \\
    quadramp\sb2  & 0.0025253 &   12.94     & 0.1224  \\
    logramp\sb1   & 0.0025267 &   14.22     & 0.0921  \\
    risingexp     & 0.0025274 &   15.04     & 0.0814  \\
    \hline
\end{tabular}\hfill\strut
\end{table}

In contrast, we discarded the initial \waspf2 light curve past the
eclipse ingress due to the disrupted ramp variation.  The eclipse
model parameters are thus less constrained.  By this point, we already
had single-channel fits for the rest of the data, so we tuned the
\waspf2 analysis in a joint fit with all the other events, sharing the
eclipse duration and mid-time.  SDNR indicates the B-subtract method
with 4.00 pixel A aperture as the best dataset.  Table
\ref{table:wa008bs42ramps} presents the four best-fitting models. The
eclipse depth is consistent with the \waspf1 depth.

\begin{table}[ht]
\caption{\label{table:wa008bs42ramps} 
\waspf2 Ramp Model Fits}
\strut\hfill\begin{tabular}{ccccc}
    \hline
    \hline
    R(t)          & SDNR      & $\Delta$BIC & Ecl. Depth (\%)\\
    \hline
    linramp       & 0.0032320 &  0.00            & 0.0932   \\
    quadramp\sb1  & 0.0032312 &  6.19            & 0.0892   \\
    logramp\sb1   & 0.0032321 &  6.67            & 0.0938   \\
    risingexp     & 0.0032330 &  7.08            & 0.0961   \\
    \hline
\end{tabular}\hfill\strut
\end{table}

\subsubsection{Final Joint-fit Analysis}
\label{secjoint}

\begin{table*}[ht]
\centering
\caption{\label{table:fits} Best-Fit Eclipse Light-Curve Parameters}
\setlength{\tabcolsep}{4.5pt}
\begin{tabular}{rccccccc}
\hline
\hline
Parameter                                       & wa008bs11       & wa008bs21     & wa008bs22     & wa008bs23     & wa008bs41     & wa008bs42    \\ 
\hline                                                                                                                                           
Array Position ($\bar x$, pix)                  & 14.74           & \n20.76      &  \n20.39      & 14.65         &  \n19.11      &  \n18.72     \\ 
Array Position ($\bar y$, pix)                  & 15.07           & 233.30        & 233.30        & 15.12         & 230.27        & 230.20       \\ 
Position Consistency\tnm a ($\delta\sb x$, pix) & 0.0072          & 0.0223        & 0.0220        & 0.0097        & 0.0273        & 0.0254       \\ 
Position Consistency\tnm a ($\delta\sb y$, pix) & 0.0118          & 0.0228        & 0.0236        & 0.0101        & 0.0272        & 0.0274       \\ 
A Aperture Size (pix)                           & 2.25            & 3.5           & 3.75          & 2.25          & 3.5           & 4.0          \\ 
WASP-8B photometric correction                  & subtract        & 1.6 mask      & 1.8 mask      & subtract      & 1.6 mask      & subtract     \\ 
System Flux $F\sb{s}$ (\micro Jy)               & 144555.0(21.0)  & 91369.9(8.5)  & 90850.3(8.5)  & 87473.0(21.0) & 32892.5(6.6)  & 34949.8(8.9) \\ 
Eclipse Depth (\%)                              & 0.113(18)       & 0.0692(68)    & 0.0692(68)    & 0.0692(68)    & 0.093(23)     & 0.093(23)    \\ 
Brightness Temperature (K)                      & 1552(85)        & 1131(35)      & 1131(35)      & 1131(35)      & 938(99)       & 938(99)      \\ 
Eclipse Mid-point (orbits)                      & 0.51428(34)     & 0.51446(37)   & 0.51468(41)   & 0.51536(28)   & 0.51446(37)   & 0.51468(41)  \\ 
Eclipse Mid-point (MJD\sb{UTC})\tnm b           & 5401.4981(28)   & 4822.2301(31) & 4814.0732(33) & 5409.6656(23) & 4822.2301(31) & 4814.0732(33)\\ 
Eclipse Mid-point (MJD\sb{TDB})\tnm b           & 5401.4989(28)   & 4822.2309(31) & 4814.0739(33) & 5409.6663(23) & 4822.2309(31) & 4814.0739(33)\\ 
Eclipse Duration ($t\sb{\rm 4-1}$, hrs)         & 2.600(78)       & 2.600(78)     & 2.600(78)     & 2.600(78)     & 2.600(78)     & 2.600(78)    \\ 
Ingress/Egress Time ($t\sb{\rm 2-1}$, hrs)      & 0.314           & 0.314         & 0.314         & 0.314         & 0.314         & 0.314        \\ 
Ramp Equation ($R(t)$)                          & quadramp$_1$    & None          & None          & logramp$_1$   & linramp       & linramp      \\ 
Ramp, Linear Term ($r\sb{1}$)                   & 0.0707(70)      & ---           & ---           & 0.000504(45)  & 0.205(22)     & 0.246(37)    \\ 
Ramp, Quadratic Term ($r\sb{2}$)                & $-$3.17(75)     & ---           & ---           & ---           & ---           & ---          \\ 
Ramp, Phase Offset ($t\sb{0}$)                  & ---             & ---           & ---           & 0.4917        & ---           & ---          \\ 
BLISS Map ($M(x,y)$)                            &  Yes            & Yes           & Yes           & Yes           &  No           &  No          \\ 
Minimum number of Points Per Bin                &   5             &  5            &  4            &  4            &  ---          &  ---         \\ 
Total Frames                                    & 64320           & 2024          & 2024          & 64320         & 1012          & 1012         \\ 
Frames Used\tnm c                               & 62203           & 1936          & 1879          & 64072         & 966           & 725          \\ 
Rejected Frames (\%)                            & 3.29            & 4.35          & 7.16          & 0.39          & 4.54          & 28.36        \\ 
Free Parameters\tnm d                           & 6               & 4             & 4             & 5             & 5             & 5            \\ 
BIC Value                                       & 80444.5         & 80444.5       & 80444.5       & 80444.5       & 80444.5       & 80444.5      \\  
SDNR                                            & 0.0053772       & 0.0036250     & 0.0035698     & 0.0073926     & 0.0030768     & 0.0032320    \\ 
Uncertainty Scaling Factor                      & 0.3075          & 1.0280        &  1.0077       &  1.0902       & 1.1187        & 1.1382       \\ 
$\beta$ correction                              & 2.4             & ---           & ---           & ---           & ---           & ---          \\ 
Photon-Limited S/N (\%)                         & 37.00           & 94.71         & 96.59         & 89.66         & 76.94         & 71.00        \\ 
\hline
\end{tabular}
\begin{minipage}[t]{0.95\linewidth}
{\bf Notes:} The values quoted in parenthes are the 1$\sigma$ uncertainties.
\tablenotetext{1}{RMS frame-to-frame position difference.}
\tablenotetext{2}{MJD = BJD - 2,450,000.}
\tablenotetext{3}{We exclude frames during instrument/telescope
  settling, for insufficient points at a given BLISS knot, and for bad
  pixels in the photometry aperture.}
\tablenotetext{4}{In the individual fits.  Joint fit had
  19 free parameters.}
\end{minipage}
\end{table*}

From the three individual fits to the 4.5 \microns\ observations we
found eclipse depths of $0.072\% \pm 0.021\%$, $0.086\% \pm 0.022\%$,
and $0.068\% \pm 0.007\%$ for \waspt1, \waspt2, and \waspt3,
respectively.  The weighted mean of the depths is $0.0692\% \pm
0.0065\%$. With a dispersion of 0.0062\% around the mean, this is not
larger than the individual uncertainties, thus we found no evidence
for temporal variability. This dispersion corresponds to 10\% of the
mean eclipse depth.  The consistency permitted a joint analysis of all
observations. We used the best light curves and models found in the
individual fits, where all events shared the eclipse duration, the
three 4.5-\micron\ events shared their eclipse depth, the two
8.0-\micron\ events shared their eclipse depth, the simultaneous
\waspt1 and \waspf1 events shared their eclipse mid-point phases, and
the \waspt2 and the \waspf2 events shared their eclipse mid-point
phases. Table \ref{table:fits} shows the light-curve modeling setup
and results.  We used these joint-fit results for the orbital and
atmospheric analysis. An electronic supplement contains the best light
curves, including centering, photometry, and the joint fit.

\section{Orbital Dynamics}
\label{sec:orbit}

WASP-8b's high eccentricity ($e = 0.31$) implies that its separation
from WASP-8A at periapsis (0.055 AU) is about half that at apoapsis.
Given the argument of periapsis ($\omega=-85.86\degree$), the
secondary eclipse nearly coincides with the periapsis.  The planet,
therefore, receives over twice as much flux at eclipse as it would if
the orbit were circular, explaining in part our high brightness
temperature (see Table \ref{table:fits}).

Secondary-eclipse times can refine estimates of $e \cos \omega$ from
RV data.  The four eclipse events occurred at an
average eclipse phase of $0.514695 \pm 0.00018$.  After subtracting a
coarse light-time correction of $2a/c = 80$ s from this average phase,
we calculated $e \cos \omega = 0.02290 \pm 0.00028$ \citep[see Equation (3)
of][]{CharbonneauEtal2005apjTrES1}.  This is consistent with
\citet{Queloz2010Wasp8}, and photometrically confirms the nonzero
eccentricity of the planet's orbit (we fit $e \cos \omega$ below
without relying on the low $e$ approximation).

The eclipse timings were combined with 130 available RV data points
and with transit data from \citet{Queloz2010Wasp8} using the method
described by \citet{Campo2011} and \citet{NymeyerEtal2011}.
Forty-eight in-transit RV points were removed due to the
Rossiter-McLaughlin effect.

Our fit presented a moderate improvement to the orbital parameters of
WASP-8b (Table \ref{tab:orbit}), except for the period.  While
\citet{Queloz2010Wasp8} used several transits to measure the period,
we used their published mid-point epoch (a single date); hence, our
period is constrained mostly by our eclipses and the RV data, and thus
have a larger uncertainty.  By themselves, the secondary eclipses have
a period of $8.158774 \pm 0.00040$ days and a midpoint epoch of BJD
$2455409.6629 \pm 0.0017$ (TDB), not significantly ([$5.9 \pm 4.3]
\tttt{-5}$ days) longer than the period found by
\citet{Queloz2010Wasp8}.  The transit and eclipse periods place a
$9.8\tttt{-5}$ \degree\,day$\sp{-1}$ ($3\sigma$) upper limit on
possible apsidal precession, nearly three orders of magnitude larger
than the theoretical expectation for tidal effects
\citep{RagozzineWolf2009apjPlanetInteriors}.

\begin{table}[ht]
\vspace{-10pt}
\centering
\caption{\label{tab:orbit} Eccentric Orbital Model}
\setlength{\tabcolsep}{4.5pt}
\begin{tabular}{rr@{\,{\pm}\,}lr@{\,{\pm}\,}l}
\hline
\hline
Parameter                                 & \mctc{This work}           & \mctc{\citet{Queloz2010Wasp8}} \\
\hline
$e \sin \omega$                      & $-0.3078$    & 0.0020   & $-0.3092$   & 0.0029  \\
$e \cos \omega$                      & $0.02219$    & 0.00046  & $ 0.023$    & 0.001   \\
$e$                                  & 0.309        & 0.002    & 0.310       & 0.0029  \\
$\omega$ (\degree)                   & $-85.00$     & 0.08     & $-85.73$    & 0.18    \\
$P$ (days)                           & 8.158719     & 0.000034 & 8.158715    & 0.000016\\
$T\sb{0}$ (MJD\sb{TDB})              & 4679.33486   & 0.00057  & 4679.33509  & 0.00050 \\
$K$ (ms\sp{-1})                      & 221.9        & 0.6      & 222.23      & 0.8     \\
$\gamma_C$ (ms\sp{-1})               & $-1\,565.9$  & 0.6      & $-1\,565.8$ & 0.21    \\
$\gamma_H$ (ms\sp{-1})               & $-1\,547.4$  & 0.4      & $-1\,548.1$ & 0.6     \\
$\dot{\gamma}$ (ms\sp{-1}yr\sp{-1})  & $58.1$       & 1.2      & $58.1$      & 1.3     \\
Reduced $\chi^2$                     & \mctc{4.1}              & \mctc{0.86}           \\
\hline
\end{tabular}
\begin{minipage}[t]{0.55\linewidth}
\end{minipage}
\end{table}

\section{Atmospheric Analysis}
\label{sec:atmosphere}

\begin{figure*}[tb]
\centering
\strut\hfill 
\includegraphics[width=0.64\textwidth, trim=0 0 0 0, clip]{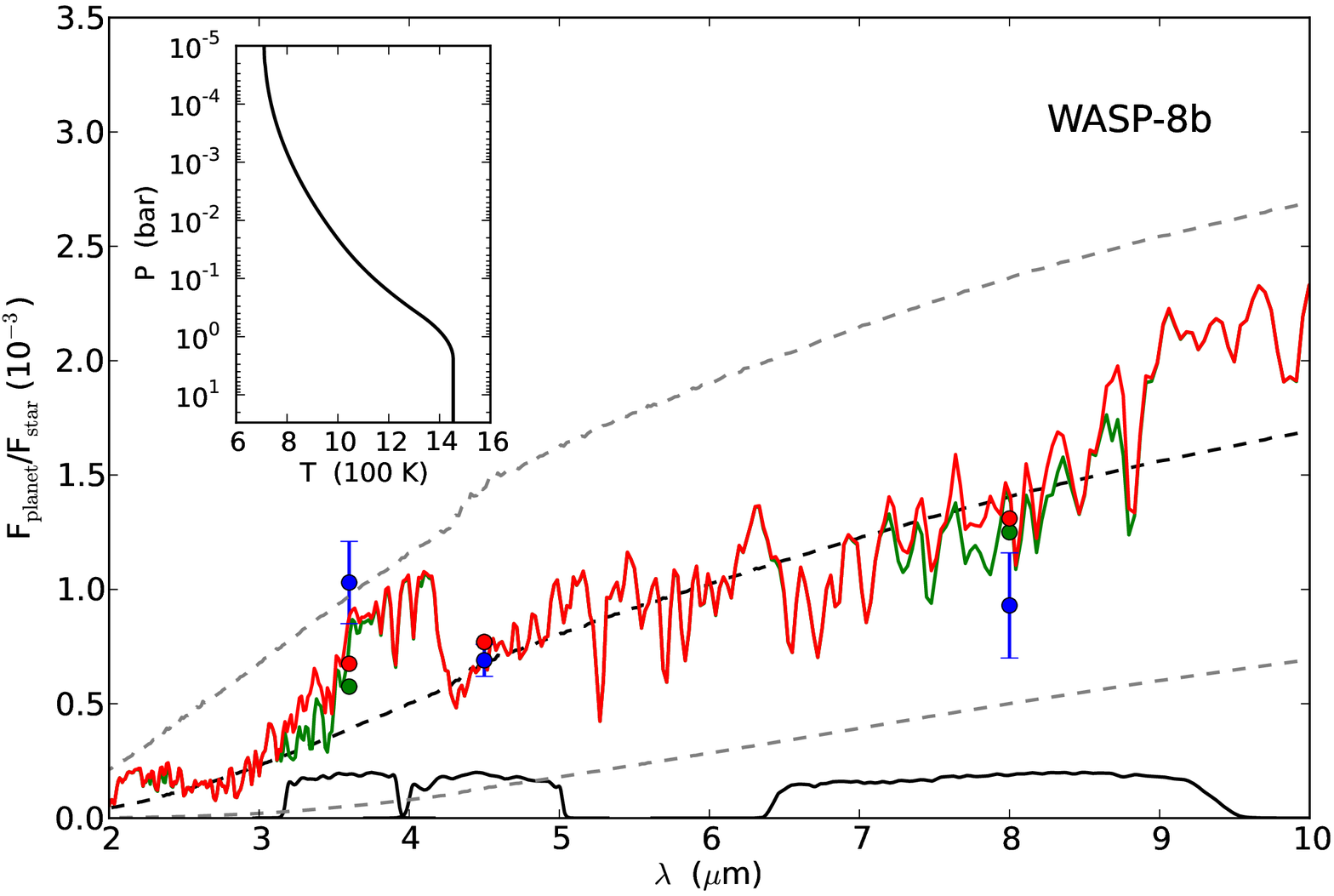}\hfill
\includegraphics[width=0.355\textwidth,trim=0 0 0 0, clip]{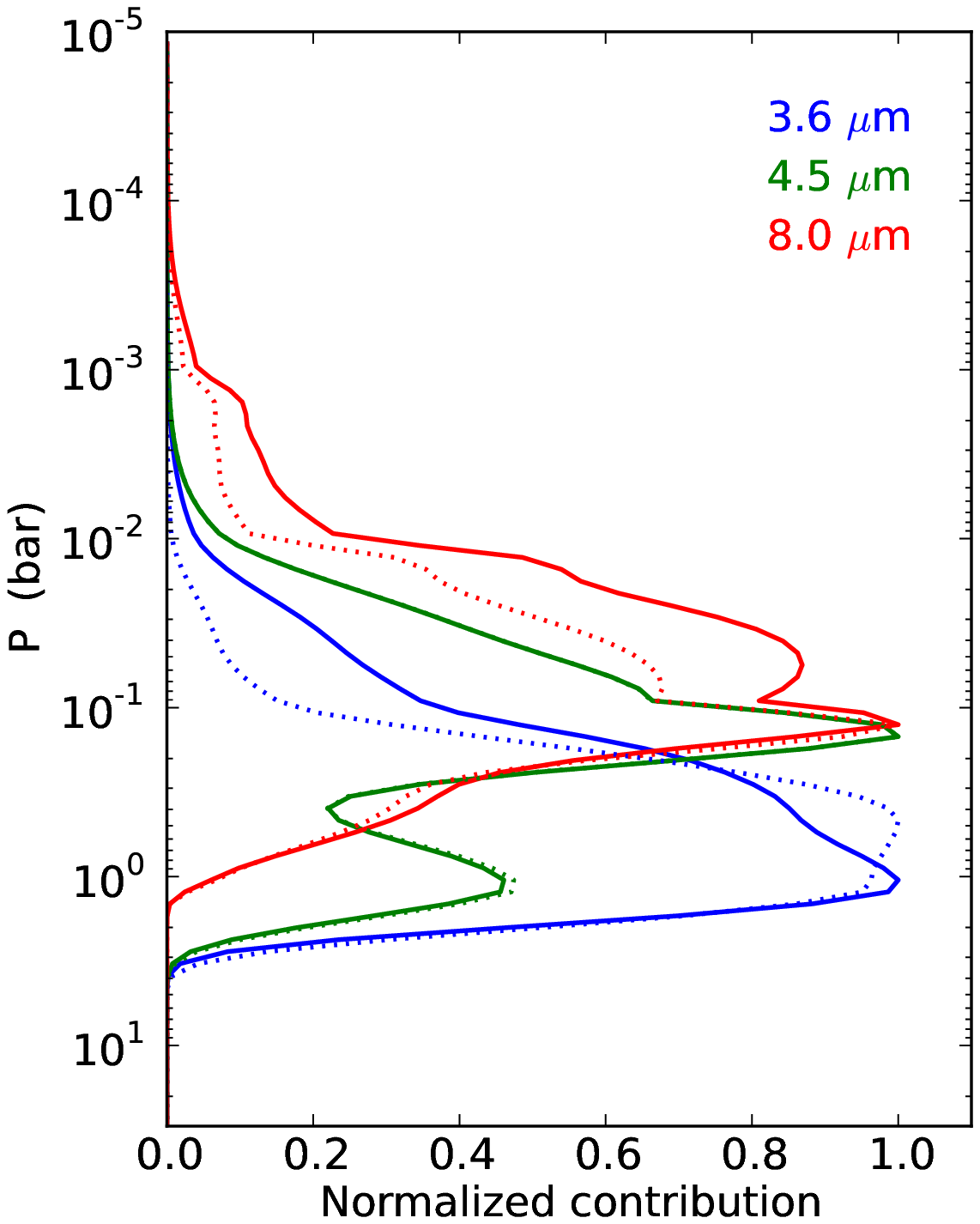}
\hfill\strut
\caption{{\bf Left:} Atmospheric spectral emission of the dayside of
  WASP-8b.  The blue circles with error bars are the measured eclipse
  depths, or equivalently, the planet-star flux ratios.  The red and
  green curves show two model spectra with the same temperature
  profile (shown in the inset) but with different compositions. The
  green curve shows a model assuming chemical equilibrium with solar
  elemental abundances. The red curve shows a model with $10\sp{3}$ times
  lower methane abundance compared to the green model, but the
  abundances of the remaining molecules are identical to those in the
  green model.  The red and green filled circles are the corresponding
  model fluxes integrated over the \Spitzer\ bands (bottom solid
  lines).  The black dashed lines represent planetary blackbody
  spectra with $T = 710, 1100$, and 1450 K.  {\bf Right:} Normalized
  contribution functions of the solar-composition model (solid lines)
  and the low-CH\sb4-abundance model (dotted lines) in each \Spitzer\
  band (see legend).  The effective pressures of the contribution
  functions are 0.63, 0.35, and 0.12 bar at 3.6, 4.5, and 8.0
  \microns, respectively.}
\label{fig:atm}
\end{figure*}

We use our IRAC observations of thermal emission from WASP-8b to
constrain the thermal structure and composition of the dayside
atmosphere of the planet.  The \Spitzer\ bandpasses at 3.6, 4.5, and
8.0 \microns\ contain strong spectral features due to several carbon
and oxygen-based molecules that are expected in hot-Jupiter
atmospheres. Methane (CH\sb4) has strong spectral features in the 3.6
and 8.0 \microns\ bands, carbon monoxide (CO) and carbon dioxide
(CO\sb2) have features at 4.5 \microns, while water vapor (H\sb2O) has
features in all three bands \citep{MadhusudhanSeager2010}. The
spectral features of the various molecules appear as absorption
troughs or emission peaks in the emergent spectrum depending on
whether the temperature decreases or increases with altitude,
respectively. Consequently, strong degeneracies exist between the
temperature structure and molecular composition derived from a
spectral dataset \citep[e.g.,][]{MadhusudhanSeager2010}. Nevertheless,
photometric observations made with \Spitzer\ have been successfully
used to constrain chemical compositions and temperature structures in
many exoplanetary atmospheres \citep[e.g.,][]{Barman2005,
  Burrows2007HD209, Knutson2008HD209, MadhusudanSeager2009apj,
  StevensonEtal2010Natur, Madhusudhan2011Nat}.

We model the dayside emergent spectrum of WASP-8b using the
atmospheric modeling and retrieval method of
\citet{MadhusudanSeager2009apj, MadhusudhanSeager2010}. The model
computes line-by-line radiative transfer in a plane-parallel
atmosphere assuming hydrostatic equilibrium, local thermodynamic
equilibrium, and global energy balance. We assume a Kurucz model for
the stellar spectrum \citep{CastelliKurucz2004} given the stellar
parameters.  The pressure-temperature (\PT) profile and molecular
mixing ratios are free parameters in the model, which can be
constrained from the data. The \PT\ profile comprises of six free
parameters and the mixing ratio of each molecular species constitutes
an additional free parameter.  Following
\citet{MadhusudanSeager2009apj}, we parametrize the mixing ratio of
each species as deviations from thermochemical equilibrium assuming
solar elemental abundances \citep{Burrows1999ChemEquilibrium}. We
include the dominant sources of opacity expected in hot Jupiter
atmospheres, namely molecular absorption due to H\sb2O, CO, CH\sb4 and
CO\sb2 (\citealp{Freedman2008Opacities}; R.S. Freedman 2009, private
communication), and H$_2$-H$_2$ collision induced absorption
\citep{Borysow2002H2H2}. We explore the model parameter space in a
Bayesian way using an MCMC sampler \citep{MadhusudhanSeager2010,
  MadhusudhanSeager2011}.  Given the limited number of observations
($N_{\rm obs}$=3), our goal is not to find a unique model fit to the
data; instead, we intend to constrain the region of atmospheric
parameter space that is allowed or ruled out by the data.

Our observations rule out a thermal inversion in the dayside
atmosphere of WASP-8b. This is evident from the planet--star flux
contrasts in the three IRAC bands at 3.6, 4.5, and 8.0 \microns. In
the presence of a thermal inversion, the planet--star flux contrasts in
the 4.5 and 8.0 \micron\ bands are both expected to be greater than
the flux contrast in the 3.6 \micron\ band \citep{Burrows2008,
  Fortney2008, MadhusudhanSeager2010}, due to spectral features of the
dominant molecules appearing as emission peaks as opposed to
absorption troughs. However, the low 4.5 and 8.0 \micron\ flux
contrasts relative to the 3.6 \micron\ contrast requires significant
absorption due to H$_2$O and CO across the spectrum, and hence the
lack of a thermal inversion in the atmosphere. Figure \ref{fig:atm}
shows model spectra of WASP-8b with no thermal inversion in the
temperature profile. The observed 4.5 and 8.0 \micron\ flux contrasts
are explained to a good level of fit by a model without a thermal
inversion and with solar abundance composition, as shown by the green
curve in Figure~\ref{fig:atm}. Our inference of the lack of a thermal
inversion in WASP-8b is independent of any assumption about chemical
composition or C/O ratio \citep[e.g.][]{MadhusudhanSeager2011}.  The
lack of a thermal inversion in WASP-8b is not surprising, since it is
amongst the cooler population of irradiated hot Jupiters, which are
not expected to host inversion-causing species such as TiO or VO in
their upper atmosphere \citep{Fortney2008, Spiegel2009TiO}.

Our models are unable to reproduce the high planet--star flux contrast
observed in the 3.6 \micron\ IRAC band, independent of the
composition. The major sources of absorption in the 3.6 \micron\ band
are H\sb2O and CH\sb4. In principle, decreasing the CH\sb4 and/or
H\sb2O abundances can lead to a higher 3.6 \micron\ contrast. However,
as shown by the red curve in Figure~\ref{fig:atm}, such an increase also
simultaneously increases the contrast in the 8.0 \micron\ band,
thereby worsening the fit overall. Another hindrance to fitting the
observed 3.6 \micron\ contrast is that it requires a hotter \PT\
profile, with $T \gtrsim$ 1550 K in the lower atmosphere, predicts
much higher fluxes in the 4.5 and 8.0 \micron\ bands than observed. On
the other hand, a cooler \PT\ profile than shown in Figure~\ref{fig:atm}
would provide a better fit in the 4.5 and 8.0 \micron\ bands, but
would further worsen the fit in the 3.6 \micron\ band.  Consequently,
we choose an intermediate \PT\ profile that provides a compromise fit
to all three data points.

Although the one-dimensional (1D) models shown in Figure~\ref{fig:atm}
output less energy than the instantaneous incident irradiation during
the eclipse (concurrent with periastron passage), they output $\sim$
20\% higher energy compared to the time-averaged incident irradiation
received at the substellar point.  Considering that a
pseudo-synchronous rotation should facilitate the redistribution of
energy to the night side, the high emission measured suggests that
WASP-8b is quickly reradiating the incident irradiation on its dayside
hemisphere, i.e. nearly zero day--night redistribution.  Such a
scenario would lead to a large day--night temperature contrast in the
planet which can be confirmed by thermal phase curves of the planet
observed using warm Spitzer
\citep[e.g.,][]{KnutsonEtal2009apjHD149026bphase}. The high emergent
flux also implies a very low albedo, as with most hot-Jupiter planets
\citep{CowanAgol2011Albedos}.

\section{The Unexpected Brightness Temperatures of WASP-8\lowcase{b}}
\label{sec:discussion}

As seen in the previous section, the 3.6-\micron\ brightness
temperature is anomalously higher than expected.  The
hemisphere-averaged equilibrium temperature for instantaneous
reradiation (time-averaged around the orbit) is only 948 K; even the
instantaneous equilibrium temperature at periapsis, 1128 K, is far
lower than this observation.  Thus, we modeled the orbital thermal
variation due to the eccentricity to determine if such a high
temperature is possible from irradiation alone.

Following \citet{CowanAgol2011TVariation}, we solved the energy
balance equation in a one-layer latitude--longitude grid over the
planetary surface.  The change in temperature of a cell with time,
${{\mathrm d} T}/{{\mathrm d} t}$, is determined by the difference
between the absorbed flux from the star and the re-emitted blackbody
flux,
\begin{equation}
\frac{{\rm d} T}{{\rm d} t} = \frac{1}{c\sb h} \left[ (1-A) \sigma T\sp4\sb{\rm eff} \left(\frac{R\sb{*}}{r(t)} \right)\sp2\cos\psi(t)\ -\ \sigma T\sp4 \right],
\label{eq:ebalance}
\end{equation}
where $c\sb h$ is the heat capacity per unit area; $T\sb{\rm eff}$ and
$R\sb*$ are the star's effective temperature and radius, respectively;
$r(t)$ is the planet--star separation; $\cos \psi(t) =
\sin\lambda\,{\rm max}(\cos\Phi(t),0)$ is the cosine of the angle
between the vectors normal to the planet surface and the incident
radiation, with $\lambda$ the latitude of the cell and $\Phi(t)$ the
longitude from the substellar meridian; $\sigma$ is the
Stefan--Boltzmann constant.

Tidal interactions drive the planet's rotational angular velocity
($\omega\sb{\rm rot}$) toward synchronization with the orbital angular
velocity ($\omega\sb{\rm orb}$).  Hence, if the spin synchronization
timescale \citep[e.g.,][]{SeagerHui2002, Goldreich1966Q} is shorter
than the system age, we expect $\omega\sb{\rm rot} =\omega\sb{\rm
  orb}$.  In the case of WASP-8b, the timescale for tidal
synchronization is on the order of 0.05 Gyr, much shorter than the age
of the star.  However, a planet in an eccentric orbit, where
$\omega\sb{\rm orb}$ changes in time, is actually expected to reach a
pseudo-synchronization state \citep[e.g.,][]{Langton2008, Hut1981}, in
which the planet does not exchange net angular momentum with its
orbit. The planet acquires then a constant rotational angular velocity
close to the orbital angular velocity at periastron ($\omega\sb{\rm
  orb,p}$). In the literature we found different predictions for this
equilibrium angular velocity, from $0.8\, \omega\sb{\rm orb,p}$
\citep{Hut1981} to $1.55\, \omega\sb{\rm orb,p}$ \citep{Ivanov2007}.

The tidal evolution drives the orbit of a planet toward zero obliquity
in a timescale similar to the spin synchronization \citep{Peale1999}.
We thus adopted zero obliquity for our simulations.  We also assumed
$A=0$, supported by the atmospheric analysis (Section
\ref{sec:atmosphere}).  Beyond these assumptions, the parameters of
interest that control Equation (\ref{eq:ebalance}) are the radiative
time $\tau_{\rm rad}=c\sb h/\sigma T\sp3\sb0$ (where $T\sb0$ is the
substellar equilibrium temperature at periastron) and the rotational
angular velocity of the planet $\omega_{\rm rot}$ (which determines
the substellar longitude of a cell through the equation ${\rm
  d}{\Phi}(t) /{\rm d}t = \omega\sb{\rm rot} - \omega\sb{\rm
  orb}(t)$).  With these definitions Equation (\ref{eq:ebalance}) can
be re-written as:
\begin{equation}
\frac{{\rm d} T}{{\rm d} t} = \frac{T\sb0}{\tau\sb{\rm rad}} \left[ \left(\frac{a(1-e)}{r(t)} \right)\sp2 {\rm max}(\cos\Phi(t),0)\  -\ \left(\frac{T}{T\sb0}\right)\sp4 \right].
\label{eq:ebalance2}
\end{equation}

We derived the temperature of each cell as a function of time to study
its thermal evolution.  Assuming that each cell emits as a blackbody,
we calculated the photometric phase curve of the planet by integrating
over the hemisphere observable from Earth, weighted by the viewing
geometry.  Our simulations were for planets nearly in
pseudo-synchronous rotation ($\omega\sb{\rm rot} =$ 0.8, 1.0, and
$1.5\ \omega\sb{\rm orb,p}$). We tested values of $\tau$\sb{rad}
between 1 and \ttt3 hours.

\begin{figure}[tb]
\centering
\includegraphics[width=\linewidth, clip, trim=20 0 47 20]{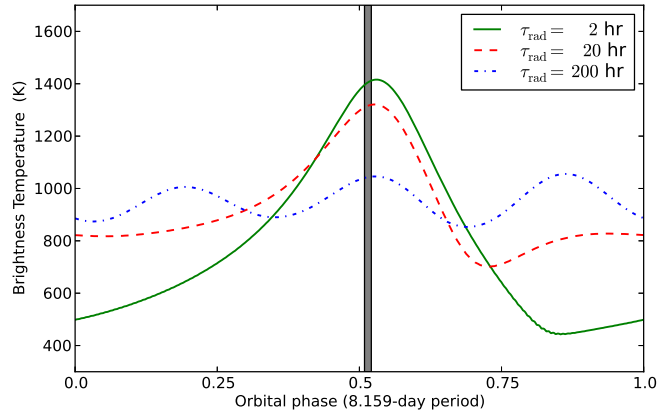}
\caption{Model brightness-temperature lightcurves of WASP-8b as observed from
  Earth during one orbital period. Phase zero indicates the mid-transit
  time. The gray region indicates the secondary-eclipse interval, with
  periastron at phase 0.52. The models simulate super-rotating winds
  ($\omega\sb{\rm rot} = 1.5\ \omega\sb{\rm orb,p}$) for different
  radiative times (see legend). The curves with smaller $\tau$\sb{rad}
  show larger amplitudes.  For $\tau$\sb{rad} comparable to the orbit
  period, and since $\omega\sb{\rm rot} = 1.5\ \omega\sb{\rm orb,p}$,
  opposite sides will face the star during successive periastron 
  passages, leading to two bright spots and hence three periodic peaks
  per orbit.}
\label{fig:model}
\end{figure}

Figure \ref{fig:model} shows simulated brightness-temperature
lightcurves of WASP-8b after reaching a periodic stationary state
(after a few $\tau$\sb{rad}).  We noted that the higher irradiation at
periastron is not the only contribution to a higher temperature.  For
$\omega\sb{\rm rot} \geq \omega\sb{\rm orb,p}$, the substellar
angular velocity (${\rm d}{\Phi} /{\rm d}t$) is minimum during
periastron, allowing the temperature to increase due to the longer
exposure to the irradiation.  For $\omega\sb{\rm rot} < \omega\sb{\rm
  orb,p}$, the substellar angular velocity is negative for an instant
around periastron. Later, when the planet emerges from secondary
eclipse the over-heated region becomes observable from Earth. As a
result, the lightcurve shows a delayed maximum.

Our models show that for large radiative timescales, the temperatures
at secondary eclipse are lower than 1150 K, regardless of
$\omega\sb{\rm rot}$.  For radiative times shorter than $\sim \ttt2$
hr, the temperatures can be as high as 1400 K, similar to the
3.6-\microns\ measurement (Figure \ref{fig:model}, top panel).
However, these models still cannot explain the observed
brightness-temperature discrepancy with wavelength.

The study of eccentric hot-Jupiter atmospheric circulation by
\citet{KatariaEtal2012CircEccentric} hints at a resolution to this
discrepancy.  Their Figure\ 4 (top panel) shows that, as the planet
passes through periapsis, the time that the peak temperature is
reached varies as a function of pressure. This is typical of their
simulations (T. Kataria 2009, private communication).  If this
differential response is significant in WASP-8b, it would introduce a
discrepancy in the observations since the \Spitzer\ bands sample
different altitudes (see Figure \ref{fig:model} right panel).

Another possibility is to compare the radiative and advective
timescales at the altitudes sampled by each band.  Evaluating equation
(1) of \citet{Fortney2008} using WASP-8b's {\PT} profile, indicates that
$\tau$\sb{rad} increases with depth between 0.1 and 1.0 bar, so there
should be less longitudinal temperature contrast at depth.  On the
other hand, models of \citet{KatariaEtal2012CircEccentric} show that
wind speeds decrease with depth, and thus $\tau$\sb{adv} also
increases with depth.  If the increase of $\tau$\sb{adv} with depth is
sharper than that of $\tau$\sb{rad}, then one would expect
less-homogenized temperatures at depth (but still above the
photosphere).  Hence, the rise in temperature (due to the increasing
incident irradiation) near periapsis could be more pronounced at 3.6
{\microns} than at longer wavelengths, given the weighting functions
of Figure \ref{fig:atm}.

\section{Conclusions}
\label{sec:conclusions}

\Spitzer\ observed secondary eclipses of WASP-8b in the 3.6, 4.5, and
8.0 \microns\ IRAC wavebands.  In our joint-fit model, we estimate
eclipse depths of $0.113\% \pm 0.018\%$, $0.069\% \pm 0.007\%$, and
$0.093\% \pm 0.023\%$ at 3.6, 4.5, and 8.0 \microns,
respectively. These depths correspond to brightness temperatures of
1552, 1131, and 938 K, respectively.  Although the 3.6-\microns\
eclipse depth is unexpectedly large, most of the ramp models had
consistent depths (within 1$\sigma$), while those with inconsistent
depths fit the data poorly.

Considering the \PT\ profile of WASP-8b, KCl, ZnS, Li\sb{$2$}, LiF, or
Na\sb{$2$}S clouds could form \citep[see Figure 2(a)
of][]{LoddersFegley2006Chem}.  In analogy to brown dwarfs, partial
cloud coverage can cause photometric variability
\citep{Artigau2009BrownDwarfs}; however, our three 4.5 \micron\
observations, spanning 1.5 years, have consistent eclipse depths,
suggesting no temporal variation at secondary eclipse above a
hemispheric-mean level of $\sim 35$ K (1$\sigma$).  A moderate cloud
layer at altitudes higher than those probed by \Spitzer\ would produce
a featureless planetary spectrum at wavelengths shorter than 2
\microns\ \citep{Pont2008HD189Haze, Miller-Ricci2012Gj1214clouds} and
would block some of the stellar flux, decreasing the temperatures at
levels probed by {\Spitzer}.  Yet, the observed temperatures, which
exceed the time-averaged equilibrium temperature, challenge this idea.

Given the high eccentricity, spin-orbit misalignment, and observed
RV drift in the of WASP-8 system, \citet{Queloz2010Wasp8}
suggested the existence of an additional, unseen body in the system.
Our orbital analysis is consistent with theirs.  It also improves the
orbital parameters and extends the baseline of sampled epochs.  This
constrains the long-term evolution of the orbit and aids the search
for a second planet, for example through the study of timing
variations \citep[e.g.,][]{AgolEtal2005TTV}.

The eclipse depths probe the dayside atmosphere of WASP-8b.  Our
results rule out the presence of a thermal inversion layer, as
expected, given the irradiation level from the host star.  A model
with solar-abundance composition explains the 4.5 and 8.0 \microns\
planet-star flux contrast; however, including the high 3.6 \microns\
flux contrast requires models that output nearly 20\% of the
orbit-averaged incident irradiation, independent of the atmospheric
composition.  If the orbit were circular (and thus the irradiation
steady-state), the high brightness temperatures would indicate a very
low energy redistribution to the night side of the planet.  For an
eccentric planet, it at least indicates a short $\tau$\sb{rad} (Figure
\ref{fig:model}).

By modeling the orbital thermal variations due to the eccentricity of
the orbit, we determined that it is possible for WASP-8b to achieve
temperatures as high as the 3.6 \microns\ brightness temperature.
However, the differing brightness temperatures in the other two bands
remain puzzling.  Neither the radiative-transfer model (Section
\ref{sec:atmosphere}) nor the phase-variation model (Section
\ref{sec:discussion}) embraces all the physics of the problem. The
radiative transfer code is a 1D, steady-state model representing
typical dayside conditions.  The phase-variation model describes
emission as a blackbody on a single-layer grid; it does not consider
absorption or emission features from the species in the atmosphere.
Clouds \citep[e.g.,][]{CushingEtal2008apjLTdwarfs}, atmospheric
dynamics \citep[e.g.,][]{Showman2009circulation}, and photochemistry
\citep[e.g.,][]{Moses2011Disequilibrium} are not directly considered
by these models.

What we can say for certain is that the assumptions of our simple
models have been violated, which is not surprising for this eccentric
planet.  While it may be possible to construct consistent, realistic
models, model uniqueness may be elusive until more and better data are
available.

Relatively few exoplanets with equilibrium temperatures below 1500~K
have been observed at secondary eclipse \citep{CowanAgol2011Albedos}.
The same is true for eccentric planets.  The characterization of
WASP-8b in this work thus addresses a particularly interesting, and
challenging, region of the exoplanet phase space.  Observation of
other planets with similar equilibrium temperatures or eccentricities
will help discover the physics that drive these unusual atmospheres.\\

\acknowledgments
We thank contributors to SciPy, Matplotlib, and the Python Programming
Language; the free and open-source community; the NASA Astrophysics
Data System; and the JPL Solar System Dynamics group for software and
services.
We thank Ian Crossfield for his help with the PSF fitting algorithm.
PC is supported by the Fulbright Program for Foreign Students.
NM acknowledges support from the Yale Center for Astronomy
    and Astrophysics through the YCAA postdoctoral Fellowship.
This work is based on observations made with the {\em Spitzer Space
  Telescope}, which is operated by the Jet Propulsion Laboratory,
California Institute of Technology, under a contract with NASA.
Partial support for this work was provided by
NASA through awards issued by JPL/Caltech
and by the Astrophysics Data Analysis Program, grant NNX13AF38G.


\end{document}